# Proximity effect model of ultra-narrow NbN strips

I. Charaev[1*], T. Silbernagel[1], B. Bachowsky[1], A. Kuzmin[1], S. Doerner[1], K. Ilin[1], A. Semenov[2], D. Roditchev[3], D. Yu. Vodolazov[4], and M. Siegel[1]

[1]*Institute of Micro- und Nanoelectronic Systems, Karlsruhe Institute of Technology (KIT), Hertzstrasse 16, 76187 Karlsruhe, Germany*
[2]*Institute of Optical Systems, German Aerospace Center (DLR), Rutherfordstrasse 2, 12489 Berlin, Germany*
[3]*Institut des Nanosciences de Paris, Université Pierre et Marie Curie-Paris 6 and CNRS-UMR 7588, 4 place Jussieu, 75252 Paris, France*
[4]*Institute of Physics of Microstructures, Russian Academy of Sciences, 603950 Nizhny Novgorod, GSP-105, Russia*

We show that narrow superconducting strips in superconducting (S) and normal (N) states are universally described by the model presenting them as lateral NSN proximity systems in which the superconducting central band is sandwiched between damaged edge-bands with suppressed superconductivity. The width of the superconducting band was experimentally determined from the value of magnetic field at which the band transits from the Meissner state to the static vortex state. Systematic experimental study of 4.9 nm thick NbN strips with widths in the interval from 50 nm to 20 μm, which are all smaller than the Pearl's length, demonstrates gradual evolution of the temperature dependence of the critical current with the change of the strip width.

## I. INTRODUCTION

Superconductivity and geometrical effects in low-dimensional structures close to the mesoscopic and quantum limits, i.e. in structures where the thickness, width and length separately or together become comparable and smaller than the coherence length ξ and magnetic field penetration depth λ, are in focus of numerous theoretical and experimental works. This interest is stimulated by the fundamental importance of the problem itself as well as by the variety of applications of thin-film superconducting nanostructures.

Primarily, dimensionality of a superconducting film on a dielectric substrate decreases with the decrease of the film thickness $d$. Dependences of the critical temperature on the thickness of superconducting films were intensively studied both experimentally and theoretically. It has been observed that in Nb [1, 2], NbN [3, 4, 5], TaN [6], TiN [7], Pb [8], Bi [9], WSi [10] and other elementary and compound superconducting films critical temperatures decrease with the decrease in their thicknesses. It is well established that for films prepared by optimized technology and having thicknesses much larger than the coherence length the transition temperature $T_C$ is independent of the thickness and equals $T_C$ of corresponding bulk specimens. When the film thickness of a uniformly disordered superconductor decreases to a value in the range from 10 to 50 nm, $T_C$ of such film starts to decrease. When the thickness further decreases to $d \leq 10$ nm, $T_C$ drops dramatically and at thicknesses less than 2 ÷ 3 nm the film transits to the insulating state (the superconductor-insulator transition - SIT). The exact values of the thickness benchmarks specified above depend on the material of superconductor, film deposition technology, and specific conditions of the film growth (crystalline mismatch to substrate material, temperature, deposition rate, partial pressure of working gases, etc.). In spite of these differences the $T_C(d)$-dependences of uniformly disordered films far from SIT in most cases are well described by the intrinsic proximity effect [11, 12]. In the framework of this model a film on the substrate is considered as a superconducting layer sandwiched between two thin layers which are either normal or possess significantly suppressed superconductivity. One of these layers is located on the surface of the film while another builds the interface between the film and the substrate. Presence of such layers has been demonstrated for NbN films [3, 13] by means of transmission electron microscopy (TEM). Thicknesses of oxidized layers were found to agree well with the thicknesses estimated in the framework of the proximity effect model. Furthermore, for NbN and NbTiN films with thicknesses of a few ξ values [14, 15], it has been shown that protecting the film with appropriate buffers and protection layers results in an increase of the critical temperature.

Secondarily, the dimensionality of a superconducting film can be reduced by patterning the film into narrow strips. In contrast to the thickness dependences, dependences of $T_C$ and other superconducting and normal state properties on the width $W$ of the strip has not been studied systematically. One of the reasons is complexity of reproducible patterning of superconducting films into elements with dimensions comparable to the coherence length of the studied material. For NbN, the coherence length is in the range from 4 to 6 nm [3] that dictates the strip widths less than one hundred nanometers. Molecular templating technique allows for fabrication of superconducting nanowires with cross-sections as small as 10×10 nm$^2$ [16]. However, continuous and controllable variation of the only width of strips over several orders of magnitude is hardly reachable with this technique.

---

[*] ilya.charaev@kit.edu

Monotonic suppression of $T_C$ with the decrease in the width of Nb and NbN strips of different thickness has been reported [17] and described by the lateral proximity effect [12]. The width of normal (degraded) edges was extracted from the nominal, geometric widths of strips in order to fit the theory to the experimental data. The width of degraded edges in thin-film strips was estimated as one nanometer for Nb and two to three nanometers for NbN. Contrarily, the nominal density of the critical current, $j_C = I_C/(W \times d)$, where $I_C$ is the measured critical current and $W$ is the geometrical, nominal width of the strip, measured in Nb and NbN strips demonstrated a more complex dependence on the width. The $j_C(W)$-dependences at 4.2 K had different shapes depending on the thickness of the strip. Monotonic increase in $j_C(W)$ for thinner (8 nm thick Nb, 5 and 8 nm thick NbN) strips changed to non-monotonic behaviour, with a maximum in $j_C(W)$ at widths between 100 nm and 200 nm for thicker (12 and 20 nm thick Nb; 10 nm thick NbN) strips (Fig. 12c,d in [17]). Very similar, non-monotonic dependence of the nominal density of the critical current on the strip width has been observed for strips from 10 nm thick TaN film on sapphire [6].

For thick strips with widths larger than the magnetic penetration depth $\lambda$ or for thin strips ($d \ll \lambda$) with widths larger than the Pearl length $\Lambda = 2\lambda^2/d$, the nominal density of the critical current is always smaller than the density of the depairing current $j_C^{dep}$. In this specific geometrical limit of very wide strips ($W \gg \Lambda, \lambda$), the density of the supercurrent is non-uniform across the strip. It piles up at edges of the strip and the strip transits into the resistive state when the local current density reaches there the depairing value. Therefore, the nominal $j_C$ of wide strips is lower than the density of the depairing current and decreases with increasing width. Contrarily, for narrow strips ($W < (10 \div 20)\Lambda$, [18]) in the absence of applied magnetic fields and large defects, it is expected that the supercurrent is uniformly distributed over the strip cross-section and thereby the nominal $j_C$ should be independent of the width and equal to the density of the depairing current.

Theoretical description of the non-monotonic experimental $j_C(W)$-dependences was reported for 12 and 20 nm thick Nb strips [17]. Authors found that the strips obey the condition $W > \Lambda$ and considered non-uniform distribution of the supercurrent as the mechanism of the decrease of $j_C$ with increasing width for $W > 200$ nm. In order to explain the observed decrease of $j_C$ at $W < 200$ nm, the authors introduced normal edges with a width an order of magnitude larger than that width estimated from the $T_C(W)$-dependences. Since other Nb and NbN strips had smaller thicknesses and were in the opposite limit, $W < \Lambda$, the theoretical analysis above was not applied to describe $j_C(W)$-dependences observed for these thinner films. T.K. Hunt [19] studied 50 nm thick Pb strips with widths from 1 to 6 μm and found a more than twofold increase of $j_C$ in narrow strips. He assumed that the increase in $j_C$ with decreasing width and saturation of $j_C$ for the narrowest strips ($W \approx 1$ μm) were due to vortex-antivortex pairs which were generated by the self-field of the current and had the size comparable to the width of the strip.

Y. Ando et al. [20] measured 100 nm-thick Nb strips with the width from 180 nm to 10 μm and obtained at 4.2 K almost one order of magnitude increase of the nominal density of critical current. Strips narrower than 180 nm were excluded from the consideration due to a strong decrease of their $T_C$ and $j_C$ values. The $j_C(W)$-dependence presented in Ref. 20 was described by the non-uniform distribution of the local density of supercurrent and the disappearance of the barrier for vortex entry when the local density of supercurrent at edges reached approximately $0.5 j_C^{dep}$.

Recently we have found qualitatively similar dependence of $j_C$ on the width for Nb strips with widths in the same range ($0.15 \div 10$ μm) but with an order of magnitude smaller thicknesses ($9 \div 15$ nm) [21]. The decrease in the width of our Nb strips resulted in the increase of $j_C(4.2$ K$)$ by a factor of 2 to 4. The factor was found larger for thicker Nb films. The difference between observations of Y. Ando et al. [20] and our results was that in our case the decrease of $j_C$ was accompanied by the decrease in $T_C$ already at $W < 400$ nm. Furthermore, we have shown that increasing the width above 400 nm flattens the temperature dependence of $j_C$ at low temperatures as compared to $j_C(T)$-curves of narrower strips. This observation is qualitatively similar to results obtained by A.Yu. Rusanov et al. [22] for Nb and MoGe thin-film strips.

A.Yu. Rusanov et al. [22] studied a few Nb strips with different thicknesses (20 - 53 nm) and widths (1 to 5 μm). They have demonstrated that the experimental $j_C(T)$-curve is reasonably described by the temperature dependence of the depairing current in the dirty limit only for the narrowest ($W = 1$ μm) strip. However, due to simultaneous variation of the thickness and the width from strip to strip, the authors were not able to draw any conclusion on the dependence of $j_C(T)$ on the width or the thickness separately. Results for 64 nm thick MoGe strips presented by A.Yu. Rusanov et al. were limited to only three strip-widths, all larger than 2 μm. Also for this material, the $j_C(T)$-dependence was well described by the dirty-limit depairing current only for the narrowest MoGe strip at relatively large temperatures. At lowest temperatures experimental points deviated systematically downwards from the theoretical curve. Deviation of experimental $j_C(T)$-curves for Nb and MoGe strips from the theoretically calculated $j_C^{dep}(T)$ curves tended to be stronger for strips with larger width. The tendency was associated with the heating by contacts at high currents as well as with vortex flow.

The $j_C(W)$-dependence for wide strips with $W \gg \Lambda$ usually agrees well with theoretical predictions, if one considers piling-up of the local current density at the strip edges. However, non-monotonic $j_C(W)$-dependences, which was observed for strips with the widths much smaller than the Pearl length, require a more detailed analysis. Furthermore, for all thin meandering strips from NbN and TaN with the widths in the interval $W = 70 \div 240$ nm ($W \ll \Lambda$) the measured density of critical current at $T = 4.2$ K was found two to three times smaller than the depairing current density [23]. Major mechanism of $j_C$ suppression in meandering strips is the current crowding [24] in the vicinity of bends. Nevertheless, even in straight 500 nm wide Nb, NbN, and TaN strips without bends the measured critical current density was found lower than $j_C^{dep}$ [25].

Decrease of $T_C$ and $j_C$ at $T = 4.2$ K with the decrease in the strip width in the sub-micrometer range were qualitatively explained by the presence of normal (degraded) edges of the strips which suppress the superconductivity via proximity effect and therefore reduce the effective superconducting cross-section of strips. However, to the best of our knowledge there have been no attempts so far to study the effect of these normal edges onto the superconducting and normal state properties of ultra-thin and ultra-narrow strips quantitatively and self-consistently. Below we report on the thorough investigation of two series of straight strips with widths varying within three orders of magnitude in the range $W < \Lambda$. The strips were made from an NbN film with the thickness $d = 4.9$ nm, which is close to the coherence length $\xi$ ($d \approx \xi \ll W$). Due to the difference in the lithography used for strip fabrication, the effective width of damaged edge-bands along the strips differs almost by a factor of two between the two series of strips. We estimate the effective width of strips, i.e. the width of the remaining superconducting core in the strip, from the value of the magnetic field needed to drive the strip from the Meissner state to the vortex state. We will show that critical temperature, critical and retrapping current densities at $T = 4.2$ K, and residual resistivity all fall on the corresponding universal dependences on this effective width. We further describe these universal dependences in the framework of the lateral NSN proximity model in which the superconducting central part of the strip is sandwiched between two edge-bands with strongly (or completely) suppressed superconductivity. Finally, we demonstrated a gradual change in the dependence of the critical current on the temperature when the strip width increases. While $j_C(T)$-dependence for ultra-narrow strips with $W < 500$ nm coincide with the temperature dependence of the dirty-limit depairing current in almost whole explored temperature range, wider strips suffer reduction of $j_C$ at low temperatures with respect to $j_C^{dep}$. This reduction appears to be larger for strips with larger widths.

## II. TECHNOLOGY OF THIN NbN FILM STRIPS

Thin NbN films were deposited on two identical 10×10 mm$^2$, one-side polished, R-plane cut sapphire substrates by reactive magnetron sputtering of a pure Nb target in an argon and nitrogen gas atmosphere. Partial pressures of argon and nitrogen were $P_{Ar} = 1.9\times10^{-3}$ mbar and $P_{N2} = 3.9\times10^{-4}$ mbar, respectively. During deposition, the substrate was placed without being thermally anchored on the surface of a copper holder, which in turn was placed on a heater plate. During film deposition the plate was kept at a temperature of 850°C. The deposition rate of NbN was 0.14 nm/s at the discharge current of 275 mA. These conditions ensure the particular film stoichiometry which results in the highest critical temperature. The film thickness $d = 4.9\pm0.2$ nm was measured by a stylus profiler. More details on deposition regimes of thin NbN films can be found in [26].

The films on both substrates were patterned in parallel into strips by the electron-beam lithography and Ar-ion milling. The electron-beam lithography was made over the PMMA 950k resist with a thickness of approximately 95 nm. The resist on one of the substrates was exposed with a dose of ≈100 µC/cm$^2$ and developed with standard developer MIBK diluted with 2-propanol. The developer removes the exposed areas of the resist as is usual for the positive-tone PMMA technology. This chip and corresponding strips on it will be referenced through the paper as the PT-chip and PT-strips. Resist on the other chip was exposed with the two orders of magnitude higher dose ≈10 mC/cm$^2$. This high dose leads to over exposing of the PMMA resist making the over exposed resist unsolvable not only in the standard developer but also in acetone, which in this case was used as developer. The developing process with acetone removes unexposed areas of the resist and leaves on the surface of the film only over exposed resist. The chip made using such invers or negative-tone PMMA technology will be referenced below as the NT-chip (NT-strips). The large difference in the exposure dose results in different profiles and hardness of the resists on two chips. This allows us to obtain two series of NbN strips with two different level of damaging of their edges and to perform systematic investigation of influence of these edges on normal and superconducting properties of NbN strips. More details on the negative- and positive-PMMA technologies can be found elsewhere [27].

On each chip there were 20 samples each containing single straight strip of NbN embedded between two large contact pads which were made from the same NbN film. Widths of the strips varied from 50 nm to 20 μm. The smallest width was limited by the smallest feature size which is reproducibly realized by both the positive and the negative process. The largest width was chosen to be smaller than the Pearl length (for details see the next Section) that ensured uniform distribution of supercurrent across all studied strips. To avoid the current crowding [24], strips were attached to contact pads via cone-like sleeves with rounding radii 5 to 15 times larger than the strip width. The nominal width $W$ was measured for all strips by the scanning electron microscopy (SEM). The SEM images of strips made by both described technologies are shown in Fig. 1a, b.

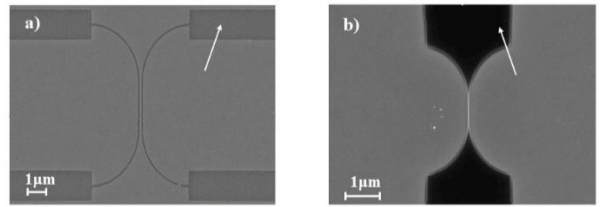

FIG. 1. SEM images of stripes which were prepared from NbN film by positive (a) and negative (b) lithography over PMMA resist. White arrows indicate the exposed areas of the resist.

## III. EXPERIMENTAL RESULTS

### A. NbN film

Temperature dependence of the film resistance was measured immediately after deposition in the temperature range from 4.2 up to 300 K by the standard four-probe technique. The resistance of the film above the superconducting transition was almost independent of temperature. The residual-resistance ratio $RRR = R(300K)/R(25K)$ was slightly larger than one, $RRR = 1.02$. The resistivity $\rho = R_{sq} \times d$ was evaluated from the measured thickness and square resistance $R_{sq}$ of the film. The residual resistivity of the film in the normal state above transition at $T = 25$ K was $\rho_{25} = 119$ μΩ×cm. The width of the superconducting transition between resistivity values of $0.1\rho_{25}$ and $0.9\rho_{25}$ was about 0.9 K. The dependence of the square resistance of the NbN film on the temperature in the vicinity of the transition (Fig. 2) is well described by the fluctuation conductivity theory of Aslamazov and Larkin [28] in the two-dimensional (2D) limit

$$R_{sq}(T) = \frac{R_{sq}}{1 + R_{sq} \cdot \frac{e^2}{16 \cdot \hbar} \cdot \frac{T_{Cf}}{T - T_{Cf}}}, \quad (1)$$

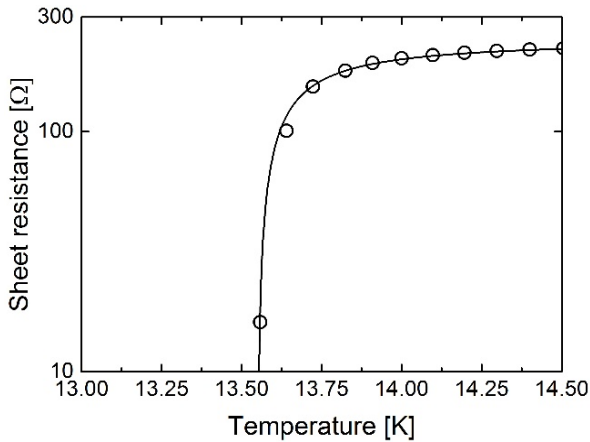

FIG. 2. Temperature dependence of the square resistance of NbN film (symbols). The solid line is the best fit by the 2D Aslamazov-Larkin fluctuation model (Eq. 1).

where $R_{sq} \approx 250$ Ohm is the square resistance of the film at $T = 25$ K and $T_{Cf} = 13.55$ K was the best fit value of the transition temperature. Temperature dependence of the second critical magnetic field $B_{C2}(T)$ was measured in the vicinity of $T_{Cf}$ in an external magnetic field applied perpendicularly to the film surface. The second critical magnetic field at $T = 0$ K was estimated from the $B_{C2}(T)$-dependence. Accounting for the dirty limit [29], we found $B_{C2}(0) = 18.8$ T and corresponding value of the coherence length at zero temperature as

$$\xi(0) = \sqrt{\frac{\Phi_0}{2\pi B_{C2}(0)}}, \quad (2)$$

where $\Phi_0$ is flux quantum. The coherence length $\xi(0) = 4.2$ nm was close to the thickness of the film. Electron diffusion coefficient $D = 0.56$ cm$^2$/sec was calculated from the linear part of the temperature dependence of $B_{C2}$ near $T_{Cf}$ with the following expression

$$D = -\frac{4k_B}{\pi e}\left(\frac{dB_{C2}}{dT}\bigg|_{T \to T_{Cf}}\right)^{-1}. \quad (3)$$

The penetration depth of magnetic field at zero temperature, $\lambda(0) = 287$ nm, was obtained with the expression

$$\lambda(0) = \sqrt{\frac{\hbar \rho_{25}}{\pi \mu_0 \Delta(0)}}, \quad (4)$$

where $\Delta(0) = 2.05\, k_B T_C$ is the superconducting energy gap for NbN films of similar thickness [30, 31]. The value of the Pearl length $\Lambda = 2\lambda^2/d = 33.2$ μm was larger than the width of the widest (20 μm) strip in our study. From the theoretical point of view, the inequality $W < \Lambda$ ensures uniform distribution of the supercurrent across all studied strips.

### B. NbN strips

#### 1. Critical temperature

Temperature dependence of the resistance for all strips was measured by a quasi-four-probe technique where the current leads and voltage contacts were made by Al-wires bonded to the large contact pads. Since pads are connected in series with the strip and contribute to the measured resistance, fitting the transition with the fluctuation theory fails to estimate the critical temperature of the strip. Instead, the critical temperature $T_C$ of the strips was defined as the temperature at which the total measured resistance drops to $10^{-3}R(25$ K$)$. For typical resistance values of our strips in excess to several kOhm, this criterion corresponded to measured voltages well above the noise level of our experimental setup at a bias current of 100 nA. The dependences of the critical temperature on the nominal strip-width $W$ for PT-chips and NT-chips are shown in Fig. 3. For strips wider than 1 μm, the critical temperatures of the PT-strips and NT-strips are almost independent on the width and are close to each other and to the $T_{Cf}$ of the non-patterned NbN film. For both series of strips with widths in the sub-micrometer range, $T_C$ decreases monotonically with the decrease in the width. It is seen that at $W < 200$ nm suppression of $T_C$ is stronger for PT-strips. Consequently, the critical temperature of the NT-strips is approximately 0.5 K higher in the range of widths from 50 nm to ≈100 nm.

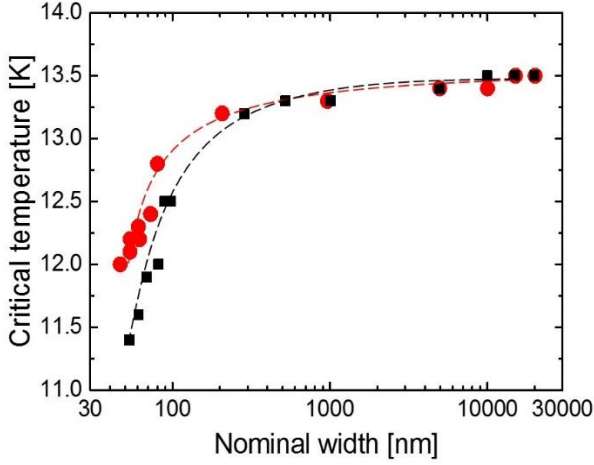

FIG. 3. Dependence of the critical temperature on the nominal width for PT-strips (black squares) and NT-strips (red circles). The dashed lines are to guide the eyes.

### 2. Critical current at 4.2 K

Current-voltage characteristics of all strips were measured at 4.2 K in the current bias mode. The critical current $I_C(4.2\ \text{K})$ of a strip was associated with a well-pronounced jump of the voltage from zero to a finite value corresponding to the resistive state of the strip. The dependences of the nominal density of the critical current, $j_C(4.2\text{K}) = I_C(4.2\text{K})/(Wd)$, on the nominal width for both series of the strips are shown in Fig. 4. Contrarily to $T_C(W)$, the $j_C(W)$-dependences are non-monotonic. When the width increases in the range from 50 up to 250 nm, $j_C$ increases too. At approximately $W = 250$ nm $j_C$ reaches the maximum value and then decreases in the micrometer range of widths. Similar dependences of $j_C$ on the width were reported for Nb [21] and TaN [6] strips with different thicknesses. The $j_C$ values of the NT-strips (red circles) are only slightly larger than $j_C$ of the PT-strips (black squares) for $W > 1$ μm. However, for the widths smaller than 1 μm the nominal density of critical current of the NT-strips is about 30% higher ($15 \div 17$ MA/cm$^2$) than $j_C$ of the PT-strips ($10 \div 14$ MA/cm$^2$).

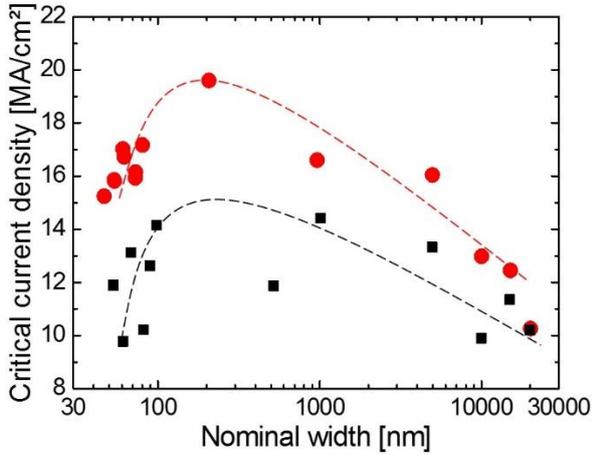

FIG. 4. Dependence of the nominal density of the critical current at 4.2 K on the nominal width for PT-stripes (black squares) and for NT-strips (red circles). The dashed lines are to guide the eyes.

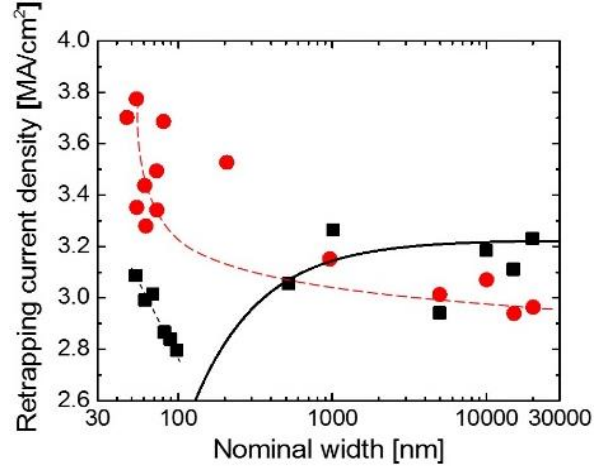

FIG. 5. Dependence of the nominal density of the retrapping current at 4.2 K on the nominal width for PT-strips (black squares) and NT-strips (red circles). The dashed lines are to guide the eyes. The solid line is the fit by Eq. 11 with $j_r^{eff}$ = 3.2 MA/cm$^2$ and $\Delta W_N$ = 25 nm as fitting parameters.

### 3. Retrapping current at 4.2 K

Retrapping current $I_r(4.2\ \text{K})$ is the current at which the strip returns from the resistive back to the superconducting state when the current decreases from a very large value to zero. This current characterizes thermal coupling of the strip to the substrate. The nominal density of the retrapping current $j_r(4.2\text{K}) = I_r(4.2\text{K})/(Wd)$ of the PT-strips shows weak dependence on the width for the micrometer wide strips (Fig. 5, black squares) and increases from 2.8 up to 3.1 MA/cm$^2$ in the range of widths smaller than 100 nm. Contrarily, $j_r$ of the NT-strips increases from 3 up to 3.8 MA/cm$^2$ with decreasing width (Fig. 5, red circles) in the whole range of the nominal width. While the nominal density of the retrapping current for micrometer wide strips is very similar for both NT- and PT-strips, for strips narrower than 100 nm the difference in $j_r$ between two types of the strips is about 20%.

## 4. Residual resistivity

The nominal residual resistivity at $T = 25$ K was calculated for each strip as $\rho_0 = R \times d / n$, where $R$ is the total resistance of and $n$ is the number of squares in the particular sample. The dependences of $\rho_0$ on the nominal width are shown in Fig. 6. The nominal residual resistivity for both series of strips grows monotonically with the decrease in the width. The residual resistivities of PT-strips and NT-strips for widths larger than 1 μm are similar to each other and are approximately 10% larger than the resistivity $\rho_{25}$ of non-patterned film. In the range of width less than 100 nm, $\rho_0$ of the PT-strips is larger than $\rho_0$ of the NT-strips. The maximal values of the nominal residual resistivity in this range of the nominal widths are 169 and 186 μΩ×cm for NT- and PT-strips, correspondingly.

## 5. Temperature dependence of the critical current

For each strip temperature dependence of the critical current was measured in the range from 4.2 K up to the $T_C$ of the particular strip. The dependence of the critical current on the normalized temperature $t = (1 - T/T_C)^{3/2}$ for the 80 nm wide strip is shown in Fig. 7a. The normalized temperature represents the temperature dependent factor in the critical current in the Ginzburg-Landau theory. There are several features in the $I_C(t)$-dependence which are typical for the strips with widths $W \leq 200$ nm (see also the black curve in Fig. 8). First of all, in the vicinity of $T_C$ ($t \rightarrow 0$) the critical current is linearly proportional to $t$ with a coefficient $I^{extr}$ (the thin blue solid line in Fig. 7a). Secondly, the experimental curve deviates from this linear growth at $t_0 \approx 0.04$ ($T/T_C \approx 0.9$), which is marked by the up arrow in the graph. In the following this normalized temperature $t_0$, the corresponding absolute temperature $T_0$, and the relative temperature $T_0/T_C$ will be referred as splitting temperatures. Thirdly, there is a narrow temperature interval located at $t \geq t_0$ where the experimental $I_C(t)$-curve transits from the linear growth at $t << t_0$ to the low-temperature regime in which the critical current further grows but with a smaller rate than neat the superconducting transition. The upper edge of this transition region at $t \approx 0.08$ ($T/T_C \approx 0.8$) is marked in Fig. 7a by the down arrow.

The shape of the temperature dependence of the critical current depends on the width of the strip. The $I_C(t)$-dependences for narrow strips with $W < 200$ nm are qualitatively similar to each other. When the width increases, the absolute splitting temperature shifts towards $T_C$ (Fig. 7b). At $W \geq 500$ nm the transition region is hardly recognizable. The $I_C(t)$-dependence of even wider strips is smooth as it is seen in Fig. 8, where $j_C(t)$-curves of strips with three different widths (54 nm, 207 nm, and 10 μm) are compared. The $I_C(t)$-curve corresponding to the narrowest strip ($W \approx 54$ nm) looks qualitatively similar to the curve for the 80 nm-wide strip presented in Fig. 7a. The $j_C(t)$-curves for the 207 nm-wide strip and for the 54 nm wide strip almost coincide near the superconducting transition of the latter strip in the range below its normalized splitting temperature $t_0 \approx 0.05$. At larger normalized temperatures, $j_C(t)$-curve for the 207 nm-wide strip deviates from the linear growth. In this region $j_C$ of the 207 nm-wide strip increases monotonically and surpasses the $j_C$ of the 54 nm wide strip. At $T = 4.2$ K $j_C$ of the 207 nm-wide strip is more than 30% higher than $j_C$ of the 54 nm-wide strip. The third curve in Fig. 8 corresponds to the 10 μm-wide NbN strip. The $j_C$ of this strip increases monotonically with the decrease in temperature. At temperatures in the vicinity of $T_C$ the $j_C(t)$-dependence of the widest strip coincides with the curves of the narrower strips. Then at temperatures below the absolute splitting temperature of the narrowest strip, $j_C$ of the 10 μm wide strip excides $j_C$ of the narrowest strip and follows the curve of the 207 nm-wide NbN strip till $t \leq 0.1$. At lower temperatures (larger $t$ values) $j_C(t)$ of the 10 μm-wide strip drops below the $j_C(t)$ of the 207 nm-wide strip. The growth rate of $j_C$ also decreases. At $t \approx 0.3$ the curve for the 10 μm-wide strip crosses the curve for the 54 nm-wide strip. Finally, at $T = 4.2$ K $j_C$ of 10 μm-wide strip is about 20% smaller than $j_C$ of the narrowest strip and 60% smaller than $j_C$ of the 207 nm-wide strip. The peculiarities in the $j_C(t)$-curves for strips with different nominal widths presented in Fig. 8 are similar for both series of NT- and PT-strips.

In summary, the $j_C(t)$-curve for 207 nm-wide strip is monotonic and stays above the corresponding curves for all other strips in the whole temperature range. $j_C(t)$-curves for strips with widths smaller or larger than approximately 200 nm move down and change the shape. $j_C(t)$-curves for strips with $W < 200$ nm reveal two distinct growing regimes (quick at $t < t_0$ and slow at $t >> t_0$) and the splitting region between them. The splitting temperature $t_0$ increases (corresponding absolute temperature decreases) with the decrease in the width. For strips with $W > 200$ nm, the grow rate of $j_C(t)$-curve at small temperatures decreases and the interval of linear grow at large temperatures shrinks.

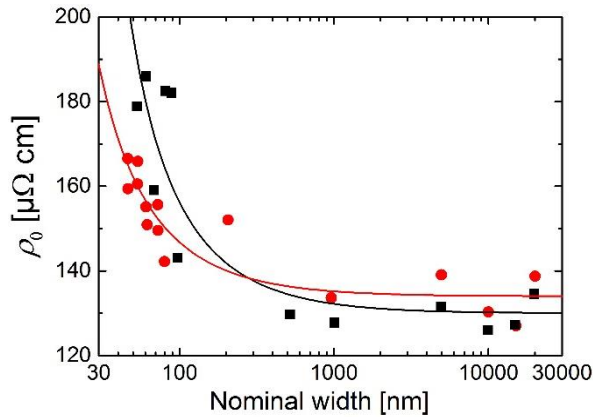

FIG. 6. Dependence of the residual resistivity $\rho_0$ on the nominal width for PT-strips (black squares) and for NT-strips. The solid lines are best fits of the experimental data by Eq. 8.

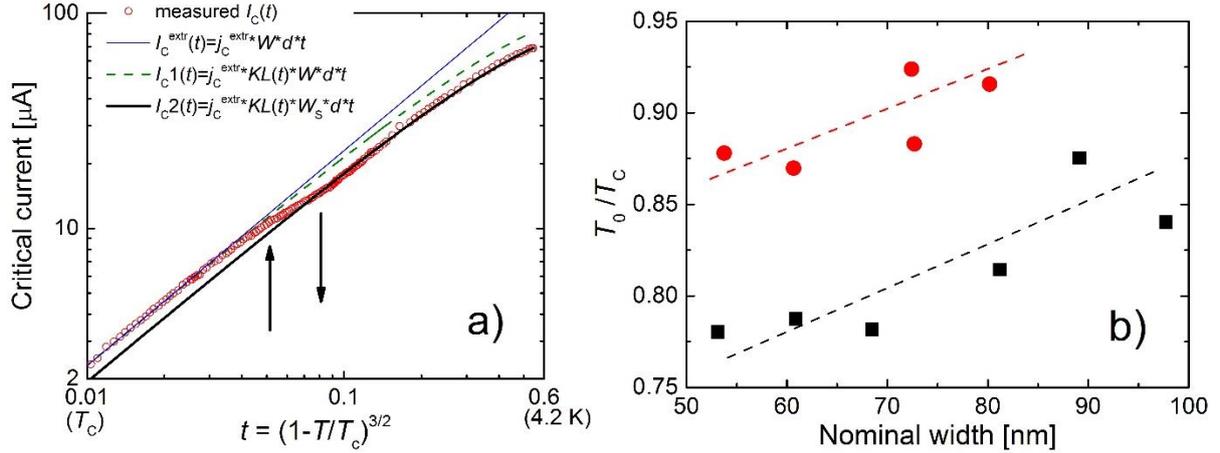

FIG.7. (a) Dependence of the critical current (red circles) for the 80 nm-wide strip (nominal width) on the normalized temperature $t = (1-T/T_C)^{3/2}$, $I_C(t)$. Thin blue solid line shows the temperature dependence of the extracted current $I_c^{extr}(t) = j_C^{extr}Wdt$ (the linear fit of the $I_C(t)$-curve at $T \to T_C$). The dashed green line ($I_C1(t)$) is the extracted current with the temperature dependent correction of Kuprijanov and Lukichev (Eq. 15) $I_c^{extr}(t)*KL(t)$. Thick black solid line ($I_C2(t)$) is the calculated critical current of the superconducting core ($W_S = 67$ nm) of the strip $j_C^{extr}*KL(t)*W_S*d$. The up arrow marks the splitting temperature $t_0$. The down arrow marks the end of the transition region. (b) Dependence of relative splitting temperature on the nominal width of PT-strips (black squares) and NT-strips (red circles).

## IV. DISCUSSION

### A. Model

We introduce here the model which allows us to explain consistently experimental results on $\rho_0$, $T_C$, $j_C$ and $j_r$ and variation of $j_C(t)$-dependences presented above in Figs. 3-8. Essentials of the model are shown schematically in Fig. 9. We assume that due to ion milling material properties of NbN are not uniform across the strip. It includes a superconducting band in the middle, which preserves properties of the original

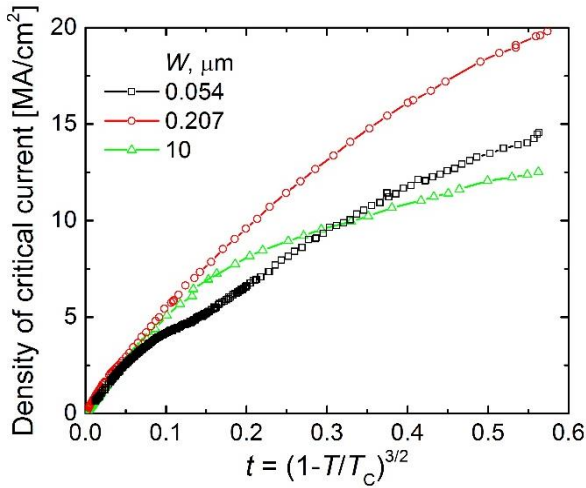

FIG. 8. Dependencies of the nominal density of critical current on the normalized temperature for three NT-strips with different widths indicated in the legend.

NbN film, and two edge-bands at both sides of the strip where superconductivity is suppressed. The suppression is the result of damaging by Ar ions during ion milling. Exact properties of the damaged edge-bands are not known. There are several scenarios of how such damaging may occur. If some parts of the strip are poorly protected by the resist etching via ion milling may decrease the thickness of the strip in those parts. The decrease of the thickness leads to suppression of superconductivity due to the intrinsic proximity effect [11, 13]. Additional suppression initiated by the milling can be even larger than the suppression expected for the film of the same final thickness prepared directly by deposition. The effect of milling is expected to be even more significant when the thickness of the original film is in the vicinity of the superconductor-insulator transition ($d \approx 2$ nm for NbN [31]). Bombardment of the film by high energy Ar ions can stimulate this transition already at larger thicknesses. Further scenario responsible for suppression of superconductivity in the edge-bands is their oxidation. After ion milling the strip is taken out of the vacuum chamber and exposed to normal air. Side walls of the strip are not protected by resist. Ion bombardment leads to "activation" of these areas by removing nitrogen atoms and liberation of Nb bonds attractive to oxygen. Bulk monoxides of Nb are superconducting at temperatures well below 4.2 K and are assumed fully normal in case of ultra-thin films. Partial oxidation reduces the strength of superconductivity in the edge bands.

We suppose that the edge-bands are normal and consider the strip as a normal-metal–superconductor–normal-metal (NSN) proximity structure, where the width of superconducting core, the effective width $W_S$, is smaller than the nominal width of the strip $W$ by the value $\Delta W_N$. We further suppose that the suppression of superconductivity in damaged edge bands is uniform across these bands (the solid frame in Fig. 9). In reality the amount of suppression is distributed (the dashed frame in Fig. 9) from the largest at the very edge till the smallest at the virtual border between the edge-bands and a superconducting core of the strip. The actual profile and the amount of suppression of superconductivity in the edge-bands is unknown but could play a crucial role in different phenomena like penetration of vortices, distribution of the order parameter and of the supercurrent across the strip.

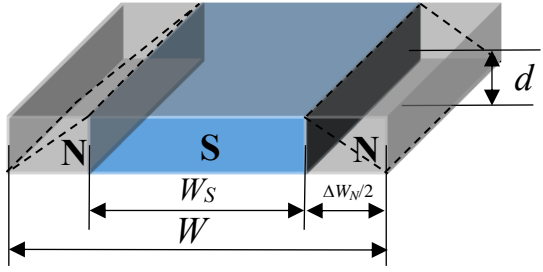

FIG. 9. Sketch of the lateral proximity system (N-S-N) in the thin superconducting NbN strip with the thickness $d$, nominal width $W$, the width of each damaged edge-band $\Delta W_N/2$, and the effective width $W_S$.

For all fabricated strips, especially those with the widths smaller than 100 nm, DC characteristics of the NT-strips correspond to a "stronger" superconductivity: higher $T_C$, $j_C$ and $j_r$ values (see Figs. 3, 4 and 5). We therefore assume that the width of damaged edge-bands ($\Delta W_N$) is smaller for the NT-strips prepared by the negative-tone PMMA lithography. The difference in $\Delta W_N$ arises from the differences in the profile and the hardness of the PMMA resist after exposure and development. In its turn, the difference in the properties of the resist is caused by significant difference in the exposure dose used in the lithographic process for NT- and PT-strips [27]. In other words, the width of the non-damaged portion of the strip, $W_S$ (the effective width), is larger in the case of strips made by the negative-tone PMMA lithography. Experimentally determined distribution of the order parameter across damaged bands with the spatial resolution comparable to the coherence length should allow for more precise analysis of experimental data and comparison with theoretical calculations of the strength of the proximity effect.

### B. Determination of the effective width

While nominal widths $W$ of strips were measured via SEM inspection, evaluation of the effective width $W_S$ required additional efforts. According to Ref. [32], the first vortex penetrates into uniform superconducting strip of the width $W_S$ at the magnetic field

$$B_{stop} = \frac{\Phi_0}{2\sqrt{3}\pi\xi(T)}\frac{1}{W_S}, \quad (5)$$

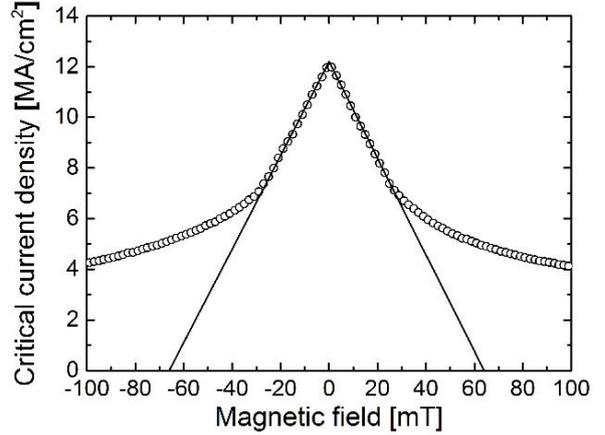

FIG. 10. Dependence of the critical current density on magnetic field for the strip with the width 1 μm at 6.2 K. Solid lines are fits of the data points corresponding to the Meissner state by Eq. 7.

where $\xi(T)$ is the temperature dependent coherence length,

$$\xi(T) = \xi(0)\left[1-\frac{T}{T_C}\right]^{-\frac{1}{2}}\left[1+\frac{T}{T_C}\right]^{-\frac{1}{4}} \quad (6)$$

and $\xi(0)$ is determined by Eq. 2.

The field $B_{stop}$ can be obtained from the dependence of the critical current on external magnetic field applied normally to the surface of the strip (see Fig. 10). The field dependence of $j_C$ has a sharp maximum and is fully symmetric with respect to the zero-field. Such behavior is expected for superconducting strips without bends and defects [33, 34]. When a straight strip without defects is in the Meissner (vortex free) state at small fields, the critical current decreases linearly with the increase in the external field $B$ as

$$I_C(B) = I_C(0)\left(1-\frac{B}{2B_{stop}}\right), \quad (7)$$

where $I_C(0)$ is the critical current in zero field.

The fit of $j_C(B)$-dependence by Eq. 7 for 1 μm wide NbN strip is shown by solid lines in Fig. 10. We used the value of $B_{stop}$ corresponding to the transition of the strip from the Meissner state to the vortex state as the only fit parameter in Eq. 7. In Fig. 10 this value corresponds to the beginning of the deviation of the $I_C(B)$-curve from the linear decrease. Using Eqs. 5 and 6, we calculated the effective widths of PT- and NT-strips from the experimentally determined $B_{stop}$ values. Calculated effective widths are shown in Fig. 11 as function of the nominal width. It is seen in the graph that effective widths of almost all strips are smaller than their nominal widths $W_S < W$. Furthermore, effective widths of the NT-strips are larger than effective widths of PT-strips. These two experimental observations confirm quantitatively the main assumptions of our NSN model. The width of one damaged edge $\Delta W_N/2$ was found to be in average less than 5 nm for the NT-strips and ≈ 15 nm for the PT-strips.

## C. Residual resistivity

Dependence of the residual resistivity on the nominal width (Fig. 6) was analyzed in the framework of our lateral NSN model. We consider a strip with damaged edge-bands as a system of two resistors with different resistances connected in parallel. The central part of the strip with the width $W_S$ contributes to the total resistance of the strip the value $R_S = \frac{\rho_S}{W_S}$, where $\rho_S$ is the normal-state resistivity of the NbN film. We further assume that the normal-state resistivity of damaged edges is larger by the value $\Delta\rho$ and amounts at $\rho_N = \rho_S + \Delta\rho$. The measurable resistivity of the entire strip is then

$$\rho_0(W) = \frac{\rho_S(\rho_S + \Delta\rho)W}{\rho_S W + \Delta\rho(W - \Delta W_N)} \quad . \tag{8}$$

At $W \gg \Delta W_N$ the resistivity of the strip $\rho_0$ does not depend on the nominal width and equals $\rho_S$. According to results presented in Fig. 11 the width of both damaged edge-bands is not larger than 40 nm. Therefore, the saturated value of the resistivity for $W > 1$ μm (Fig. 6) can be assigned to $\rho_S$ solely. The main influence on $\rho_0$ from the damaged edge-bands is expected when their width $\Delta W_N$ is comparable to $W$. Indeed, the strongest increase of resistivity is observed for $W < 100$ nm (Fig. 6). We additionally assumed that the strength of damaging of NbN film at the edge-bands and the correspondent increase of resistivity ($\Delta\rho$) are the same for both used lithographic techniques. Based on the above mentioned assumptions we fit both $\rho_0(W)$ dependences by Eq. 8 (the solid lines in Fig. 6). The best fit was obtained for $\Delta\rho \approx 160$ μΩ cm. This means that the normal-state resistivity of damaged edges is approximately twice the normal-state resistivity of the superconducting core of the strip, $\rho_S \approx 130$ μΩ cm. We neglected variations of $\Delta W_N$ for strips within PT and NT series and used for fitting single values of 30 nm and 16 nm for the PT- and NT-strips, correspondingly. These best fit values are very close to the averaged values of $\Delta W_N$ for PT- and NT-strips (25 and 10 nm) shown by the dashed lines in the inset in Fig. 11. It is seen in Fig. 6 that the fit describes the data for the NT-strips pretty good while the PT-strips show steeper increase of the residual resistivity. The discrepancy between the fit and the experimentally obtained resistivity of the PT-strips can be caused by our simplifying assumptions which rule out the possibility that $\Delta\rho$ may depend on $W$ and the type of lithographic process.

## D. Density of critical current

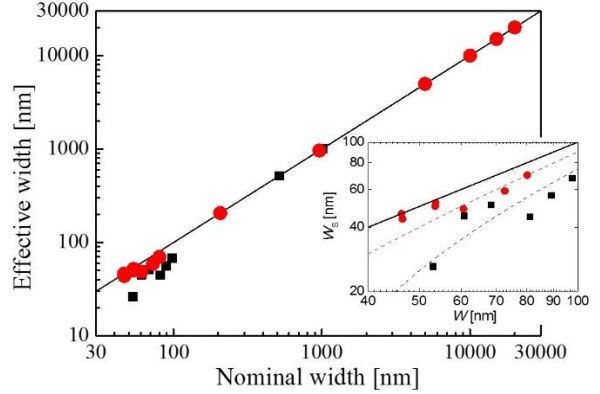

FIG. 11. Dependence of the effective width $W_S$ on the nominal width $W$ calculated with Eq. 5 for PT-strips (black squares) and NT-strips (red circles). Solid line corresponds to $W_S = W$. The inset zooms in the interval $W < 100$ nm. Dashed curves are fits by $W_S = W - \Delta W_N$: red curve - $\Delta W_N = 10$ nm; black curve - $\Delta W_N = 25$ nm

One of the essentials of the NSN model is the concentration of the supercurrent in the central part of the strip when the supercurrent is close to the measured critical current. This is true even if the edges are not completely normal. Let us suppose that the damaged edge-bands retain superconductivity but the energy gap there $\Delta^*$ is less than the energy gap $\Delta$ in the central part of the strip. At low temperatures ($T \ll T_C$) applied current $I \ll I_C$ is distributed uniformly across the strip if it is additionally smaller than the critical current corresponding to the energy gap in the edge bands. Once the density of the applied current reaches the critical value corresponding to the energy gap $\Delta^*$, the edge-bands switch into the normal state. The strip splits into parallel connected normal and superconducting parts and the supercurrent is squeezed into the central superconducting part of strip with a larger value of the energy gap. Electrically the strip remains superconducting. The transition into the normal state occurs at a larger current which corresponds to the critical current of the central band with the local density $j_C^{\text{eff}}$. The latter is determined by the energy gap $\Delta$ in the central band. Smaller $W_S$ results in a lower measured critical current and thus in a lower calculated nominal density of critical current in strips $j_C = j_C^{\text{eff}} W_S/W$. Because of smaller $W_S$ in the PT- strips, $j_C$ in these strips is lower than in the NT-strips. The difference is more pronounced in sub-micrometer range of widths as it is seen in experimental data shown in Fig. 4.

## E. Retrapping current

For $W < 100$ nm the experimentally measured nominal density of the retrapping current $j_r$ is approximately 20% larger in the NT-strips than in the PT-strips (Fig. 5). The density of the retrapping current characterizes thermal coupling between the strip and the substrate and depends on the resistivity of superconductor in the normal state and on the difference between $T_C$ and the bath temperature $T_b$ at which the current-voltage characteristic is measured [35]:

$$j_r = \left[\frac{\alpha}{4d\rho_0}\left(T_C^4 - T_b^4\right)\right]^{1/2}. \quad (9)$$

Here $\alpha$ is the coefficient characterizing thermal conductivity of the interface between the strip and the substrate. This coefficient does not depend on the strip width. Taking into account larger $T_C$ (Fig. 3) and smaller $\rho_0$ (Fig. 6) of the NT-strips in sub-100 nm range of widths, we obtained with Eq. 9 only 9% difference in the retrapping current densities in the NT- and PT-strips. This is less than one half of the experimental difference between retrapping currents in those strips.

The inconsistence between measured and estimated differences in the retrapping current finds qualitative explanation in the framework of the NSN model. Relying on the successful fit of the measured resistivity as function of the strip-width, we suggest that at $T \geq T_C$ the resistivity of the edge-bands is temperature independent and that the resistivity of edge-bands is higher than the normal-state resistivity of the superconducting core. The difference in resistivity results in a non-uniform distribution of applied current even when the strip is in the normal state at temperatures well above $T_C$. It can be shown that the ratio $i$ of currents flowing through the central band of strip and through damaged edge-bands is

$$i = \frac{1}{\frac{\rho_S \cdot \Delta W_N}{\rho_N \cdot W_S} + 1}. \quad (10)$$

With the best fit values of $\rho_S$, $\rho_N$, $W_S$ and $\Delta W_N$ (see Fig. 6 and Eq. 8) we conclude that at currents close but larger than the retrapping current even for narrowest strips more than 80% of the applied current flows through the central part of the strip. Hence, the nominal retrapping current density is to a large extent determined by $T_C$ and $\rho_S$ of the central part and the thermal interface between the superconducting core and the substrate. The retrapping current density inherent in the superconducting core $j_r^{eff}$ should be independent on the value of $W_S = W - \Delta W_N$. Thereby the nominal density of the retrapping current

$$j_r = \frac{j_r^{eff} \cdot W_S}{W} \quad (11)$$

is proportional to the width of the superconducting core, i.e. $j_r$ is larger for the NT-strips in agreement with our experimental observations. However, according to Eq. 11 the nominal density of the retrapping current should decrease with the decrease in the nominal width (the solid line in Fig. 5) if one assumes that the width of damaged edges is independent of the strip-width. Experimental data do not support this conclusion. Nominal density of the retrapping currents increases with the decrease of $W$ in sub-100 nm range of widths for PT-strips and in the whole range of widths for NT-strips. We discuss this discrepancy in the following Section.

### F. Key properties as functions of the effective width

In Fig. 12a, the critical temperature of strips is plotted as function of the experimentally determined effective width $W_S$. Experimental dependences of $T_C$ for both series of NbN strips fall into one universal curve in the whole range of the effective superconducting widths. The solid line in Fig. 12a is the best fit of the experimental data with the dependence of the critical temperature for NS proximity structure [36]

$$T_C = T_{C0}\left(1 - \left(\frac{\pi\xi_{GL}}{W_S}\right)^2\right), \quad (12)$$

which was obtained for the limiting case when the resistivity of the normal layer is much smaller than the resistivity of the superconducting layer. For the fit we used $T_{C0} = 13.48$ K and $\xi_{GL} = 4.2$ nm. Increase of the resistivity of the normal layer shifts the decay of $T_C$ to smaller values of $W_S$. Result of calculations made in the framework of the Usadel's formalism for the ratio $\rho_S/\rho_N = 0.5$ is shown in Fig. 12a by green triangles. Both theoretical approaches describe our experimental results qualitatively well. However, the latter theory predicts a weaker suppression of $T_C$ for our strips for which the resistivity of the edges is larger than the resistivity of the superconducting core. In the framework of the Usadel's formalism, it has been also found that the suppression of $T_C$ becomes independent of the width of normal edges if $\Delta W_N$ is larger than approximately $1.6\xi_{GL}$. For our NbN film, this corresponds to 7 nm. This value is close to the width of damaged edges which was obtained experimentally. Experimentally observed independence of $T_C(W_S)$ (Fig. 12a) of the width of the normal edge-bands agrees with the theoretical predictions. Universal nature of the effective width as the variable is also seen in Fig. 12b and 12c where the effective densities of the critical $j_C^{eff}$ and retrapping $j_r^{eff}$ currents ($j_{C,r}^{eff} = I_{C,r}/(W_S d)$) are plotted as functions of $W_S$. The values of $j_C^{eff}$ and $j_r^{eff}$ for both series of NbN strips fall into universal curves as it was found for values of the critical temperature. These findings support the main conclusion of the qualitative analysis made in the framework of our NSN model, i.e. transition temperatures and densities of effective currents are controlled by the properties of the superconducting core and conditions of the current transport through the central band with the width equal to the experimental effective width. In the whole range of widths, effective density of the critical current increases monotonically from 10 up to 25 MA/cm$^2$ (Fig. 12b). We now compare the effective densities obtained in the experiment with the density of the depairing current. Temperature dependence of the density of the depairing current, $j_C^{dep}$, was adopted from the publication of Kupriyanov and Lukichev [37]. We used the temperature dependent correction factor $KL(T)$ to the Ginzburg-Landau (GL) temperature dependence of the depairing current density in the extreme dirty limit and calculated $j_C^{dep}$ for our film as:

$$j_C^{dep}(T) = j_C^{GL}(0)\left(1 - \frac{T}{T_C}\right)^{\frac{3}{2}} KL(T), \qquad (13)$$

where

$$j_C^{GL}(0) = \frac{16\sqrt{\pi}\exp(2\gamma)}{21\varsigma(3)\sqrt{6}} \beta_0^2 \frac{(k_B T_C)^{\frac{3}{2}}}{e\rho_0\sqrt{D\hbar}}, \qquad (14)$$

$$KL(T) \approx 1.87 - \frac{1.46}{1 + \left(0.7\frac{T}{T_C}\right)^{1.12}} \qquad (15)$$

is the analytical approximation of the curve from Fig. 1 in Ref. 37 corresponding to the limit $\xi_{GL} \gg \ell$. In Eq. (14) $e$, $\hbar$, $k_B$ are fundamental constants, $\gamma = 0.577$, $\zeta(3) = 1.202$, $\beta_0 = \Delta(0)/(k_B T_C) = 2.05$ [30, 31] and $D = 0.56$ cm$^2$/s is the electron diffusivity in NbN film (Eq. 3). The dashed horizontal line in Fig. 12b corresponds to the value of the depairing current density, which was computed by Eq. 13 for $T_C$ and $\rho_0$ values of the superconducting core (saturated values of the critical temperature and the residual resistivity for micrometer wide strips (Fig. 6 and 12a)). It is seen that the experimental points approach the theoretical limit of the current carrying ability of the NbN film but still remain below this limit.

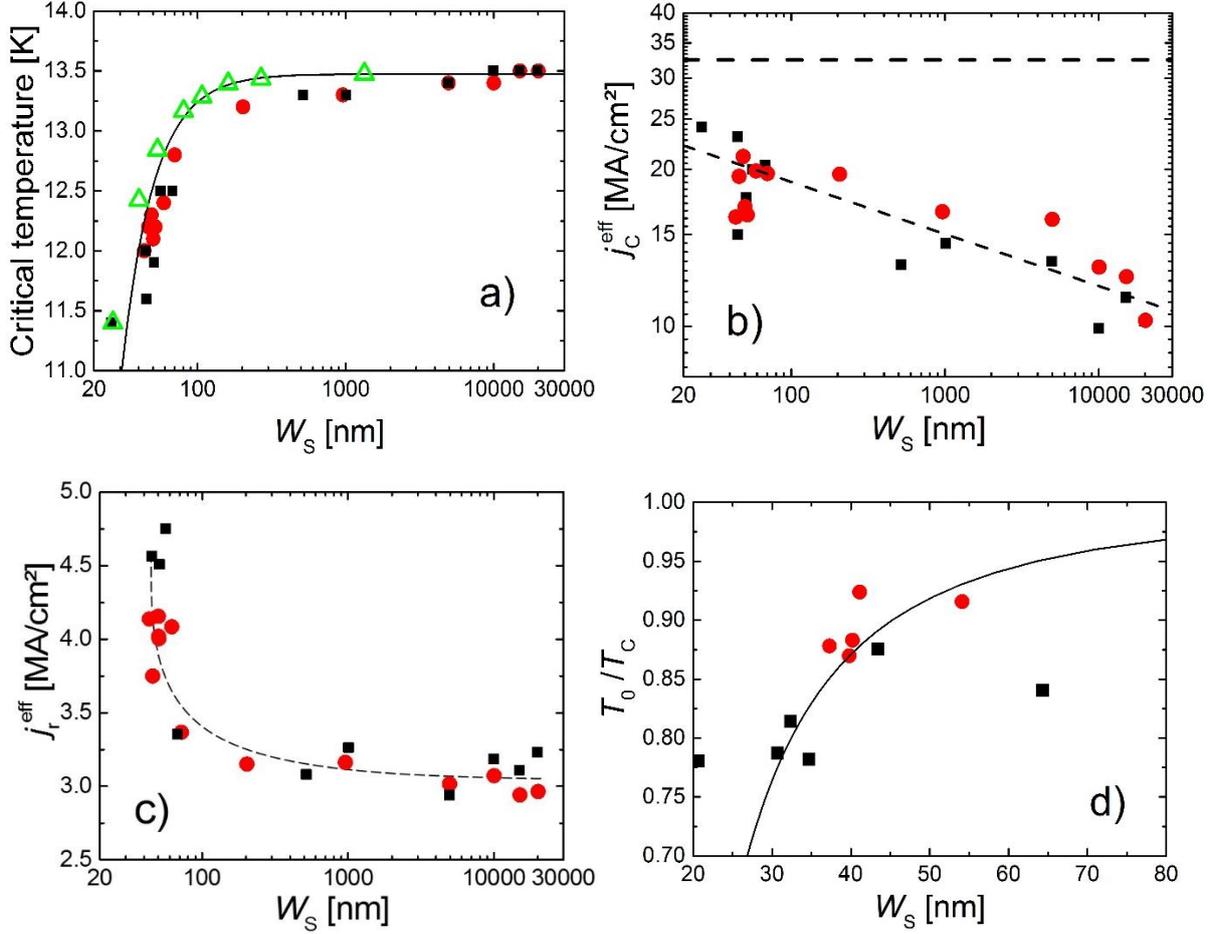

FIG.12. Dependences of the critical temperature $T_c$ (a), the effective densities of the critical current $j_C^{eff} = I_c/dW_{SC}$ (b), the effective density of the retrapping current $j_r^{eff} = I_r/dW_{SC}$ (c) currents, and the relative temperature $T_0/T_C$ of splitting (d) on the effective width of strips. (a) Solid line is the $T_C(W_S)$ dependence calculated by Eq. 12; green triangles show the same dependence computed in the framework of the Usadel's formalism for the resistivity of edge-bands twice as large as the resistivity of the central core. (b) Thick dashed horizontal line corresponds to the value of the depairing current density calculated with superconducting and normal state parameters of the superconducting core by Eq. 13. Thin dashed curves in (b) and (c) are to guide the eyes. (d) Solid curve corresponds to the inverse dependence for $W_S = 4\xi_{GL}(T_0/T_C)$ calculated by Eq. 6.

The effective density of retrapping current (Fig. 12c) is almost independent of the width at $W_S \geq 200$ nm. When the width decreases below this value, the effective density monotonically increases from 3.2 MA/cm$^2$ up to about 4.7 MA/cm$^2$. Mechanisms of the increase in $j_r^{eff}$ are unclear. One of the possible scenarios is the following. As we discussed above, due to large value and weak temperature dependence of the resistivity of the damaged edge the major part of the applied current flows through the central part of strip. Consequently, the Joule heating occurs predominantly in this central part and increases its actual temperature with respect to temperatures of edge-bands which are less heated by the current. The temperature gradient between the center of the strip and its edges on a distance approximately equal to the thermal healing length $L_h = (D\tau_\varepsilon)^{0.5}$, where $\tau_\varepsilon$ is the electron energy relaxation time, results in a lateral heat flow from "hot" center to "cold" edges. This lateral heat flow represents a channel of electron cooling additional to the channel provided by phonon escape into the substrate. When the width of the strip is much larger than $L_h$ this additional channel of electron cooling can be neglected. However, when the width of the central part ($W_S$ in our NSN model) is comparable to $2L_h$, contribution of this additional channel of the heat outflow decreases the actual temperature of the central band and, hence increases the retrapping current. For NbN the $\tau_\varepsilon$ value is in the interval from 10 to 30 ps [38]. With the diffusion coefficient $D = 0.56$ cm$^2$/s (Eq. 3), the characteristic value $W_S = 2L_h$ is between 40 and 80 nm. At this effective width one should expect an increase in $j_r^{eff}$. Our experimental results agree quantitatively with this expectation, namely $j_r^{eff}$ increases in strips narrower than 100 nm (Fig. 12c).

The width of all our strips is smaller than the Pearl length. The supercurrent is expected to be uniformly distributed over the cross-section for all strips and to be determined by depairing of Cooper pairs. In spite of the fact that for $W \leq 200$ the above condition becomes stronger $W << \Lambda$, experimental temperature dependences of the critical current density for such strips are non-monotonic (see Fig. 7a and the black curve in Fig. 8). It is not possible to describe $I_C(T)$-dependences for these strips in the whole temperature range by the temperature dependence of the depairing current. Thin blue solid line in Fig. 7a is the linear fit of the critical current at $t \to t_0$, i.e. at $T \to T_C$, by $I_C^{extr}(t) = I_C^{extr}(0)t$, where $t = (1-T/T_C)^{3/2}$ is the temperature dependence of the Ginzburg-Landau critical current. At absolute temperatures less than the splitting temperature ($t > t_0$) the experimental points deviate from the $I_C^{extr}(t)$-curve. This is not surprising from the first glance since the Ginzburg-Landau description is only valid in a very narrow temperature interval in the vicinity of the critical temperature. Correction to the Ginzburg-Landau temperature dependence suggested for the dirty limit by Kuprijanov and Lukichev, $KL(t)$, [37] allows for description of the depairing current in the whole temperature range (Eq. 13 and 15). Nevertheless, even with the $KL(t)$-correction the calculated temperature dependence $I_C1(t) = I_C^{extr}(0)KL(t)t$ (the dashed green curve in Fig. 7a) is still above the experimental data at $t \geq t_0$. Although the fit for the whole temperature range fails, it is especially noteworthy that at $t \geq 0.1$ the experimental points are offset evenly with respect to the $I_C1(t)$-curve. This low temperature part of the experimental curve can be described up to the largest value of the normalized temperature $t \approx 0.6$ (corresponds to the lowest temperature of our measurements $T = 4.2$ K) in the framework of our NSN model with the assumption that the supercurrent in this $t$-range is carried by the central superconducting band of the strip while the damaged edge-bands are excluded from the transport completely. The thick black solid curve in Fig. 7a is calculated following this assumption as $I_C2(t) = I_C1(t)W_S/W$ with $W_S = 67$ nm (see the inset in Fig. 11).

Qualitatively the NSN model is similar to the model describing a defect at the edge of the stip. According to Ref. [24] (see Fig. 18 and 19 there), strength of current crowding and the suppression of $j_C$ caused by a semicircular notch of radius $a$ both depend on the ratio $\xi/a$. While $\xi/a >> 1$, the factor of $j_C$ reduction ($R$-factor) is almost 1, i.e. the critical current is not suppressed and equals the depairing current. When the ratio decreases, the $R$-factor decreases too. At $\xi/a << 1$ the suppression is the strongest ($R$-factor reaches the smallest value). At a fixed value of the notch radius, the transition of the ratio $\xi/a$ from "much-larger" to "much-smaller" than one is governed by the temperature dependence of the coherence length $\xi(T)$. Roughly speaking, the transition of $j_C(T)$ from the undisturbed value at $T \to T_C$ to the smaller value at $T << T_C$ should occur when $\xi(T)$ approximately equals $a$. Smaller $j_C(4.2$ K) of the PT-strips (Fig. 4) would correspond to a smaller $R$-factor which for the same $\xi(4.2$ K) can be explained by a larger size of the defect. Moreover, the decrease of the strip width in the range $W < 100$ nm at a fixed size of the defect should lead to a stronger current crowding and to a stronger suppression of $j_C$. That is what we observed in our experiments (Fig. 4). However, a larger value of $a$ means that the drop of $j_C$ should occur at a larger value of $\xi(T)$, i.e. at a temperature which is closer to $T_C$. This prediction contradicts to our experimental observations in which the drop of $j_C$ in PT-strips occurs at smaller temperatures than it occurs in NT-strips (Fig. 7b).

As it is seen in Fig. 12d the dependence of the relative splitting temperature $T_0/T_C$ on the effective width follows (with the exception of two experimental points) the inverse temperature dependence of the coherence length (Eq. 6), i.e. $W_S = C\xi(T_0/T_C)$, where the prefactor C = 4 is close to the Likharev's criterion ($\approx 4.4$) for the vortex free superconducting strip [39]. In strips with the width less than $\approx 4.4\xi(T)$, energy needed for thermal activation of the phase slip line (PSL) or vortex (at $I < I_C$) are the same while in wider strips the energy for activation of vortices becomes smaller [40]. Vortex or PSL could be created in superconducting strip due to fluctuations which provide switching of the superconductor to the resistive state [41]. Therefore, the behaviour of $I_C(T)$ at $T_0$ can be interpreted (at least qualitatively) by the transition from quasi-1D (PSL) to 2D (vortex) fluctuation assisted transition of the current-carrying superconducting strip to the resistive state.

For strips wider than 200 nm, the $j_C(t)$ dependences contain no steps and are flatter as the dependences for narrow strips. The flattening with a trend to saturation becomes more pronounced when the strip-width increases. Particularly this is seen as deviation of the $j_C(t)$ curve for 10 μm-wide strip from the curve corresponding to the 207 nm-wide strip (Fig. 8). When the width increases the temperature, which corresponds to the deviation, increases too. At temperatures smaller than this characteristic temperature, $j_C(t)$ curves become flatter. Consequently, the difference between the $j_C(4.2\ \text{K})$ for the 207 nm-wide strip and $j_C(4.2\ \text{K})$ for strips in the micrometer range of widths increases with the width (Fig. 12b).

One of the mechanisms of the transition of a current-carrying superconducting strip into the resistive state is penetration and movement of vortices which are generated by magnetic field of the applied transport current. The fields corresponding to the critical currents (Fig. 12b) at the edges of the superconducting core [42]

$$B_{\text{edge}} = \frac{\mu_0 I_C \ln \frac{2W_S}{d}}{2\pi W_S} \quad (16)$$

is smaller than the experimentally measured values of $B_{\text{stop}}$ (Eq. 5). Even in the case of the widest (20 μm) strip, $B_{\text{stop}}$ is twice as large as $B_{\text{edge}}$. In the case of narrower strips with widths around 1 μm, the difference between $B_{\text{stop}}$ and $B_{\text{edge}}$ is about two orders of magnitude. Therefore, we rule out the self-field generation of the critical state in thin and narrow NbN strips. The mechanisms of the flattening of temperature dependences of the critical current in relatively wide strips ($W > 200$ nm) are under discussion.

## V. CONCLUSIONS

We have performed thorough systematic investigation of superconducting and normal-state properties of ultra-thin NbN strips with widths varying by three orders of magnitude in the range from approximately 50 nm up to 20 μm. Due to the difference in utilized lithographic techniques, strips with the same nominal widths have damaged edge-bands with different width. The width of damaged bands was found for each strip from the analysis of the dependence of the critical current on the external magnetic field which was applied normally to the surface of the strips. We have found that the width of damaged edge-bands in strips prepared by the negative-tone PMMA lithography is approximately twice as small as in the strips prepared by the standard electron–beam lithography over the positive-tone PMMA resist. In narrow strips, larger damaged edge-bands result in stronger reduction of the critical temperature at small widths, smaller densities of the critical current and the retrapping current, and larger residual resistivity. The effect is more pronounced in ultra-narrow strips with widths less than 100 nm. The observed differences between strips prepared by the negative and the positive PMMA lithography are self-consistently explained in the framework of the NSN model which implies the lateral proximity effect between the central superconducting band in the strip and two normal edge-bands as well as splitting of the current between the core and the bands. When plotted as functions of the width of the superconducting core, the dependences of the critical temperature and the densities of the critical current and the retrapping current fall into corresponding universal curves which do not depend on the width of damaged edges. The decrease of the width of the superconducting core below 100 nm results in the increase of the density of the retrapping current in the core due to opening of the diffusion channel of heat outflow from the core to edge-bands.

Besides phenomena understood in the framework of the NSN proximity model, there are a few still open questions requiring further theoretical and experimental analysis. (i) Reduction of the critical temperature in narrow strips is larger than the NSN proximity model predicts. (ii) For strips with widths much smaller than the Pearl length, the density of the critical current in the core increases with the width of the core. The self-field produced by the critical current at the edges of the core is not sufficient to generate vortices and hence does not explain variations of the critical current with the width of the core. (iii) Non-monotonic temperature dependences of the critical current are interpreted as the result of splitting of the strip into normal edge-bands and superconducting core at a characteristic temperature depending on the ratio between the widths of the core and edge-bands. The splitting causes squeezing of the current into the core and reduction of the current density in edge-bands. Although temperature dependences of the current density in these two regimes follow the temperature dependences of the depairing current, the profile of the local current density across the strip is not predicted with the NSN model. (iv) For strips with widths larger than a few hundreds of nanometers, the temperature dependences of the critical current flatten. The flattening is more pronounced and starts at higher temperatures for wider strips. The mechanism responsible for flattening is not clear. However, it is not connected with the self-field of the current.

### ACKNOWLEDGMENT

I.C. acknowledges support from Karlsruhe School of Optics and Photonics of Karlsruhe Institute of Technology. The work was supported in part by the DFG/ANR project SUPERSTRIPES SI 704/11-1. D. Yu. V. acknowledges support from Russian Scientific Foundation (grant № 17-72-30036).